# Class dependency based learning using Bi-LSTM coupled with the transfer learning of VGG16 for the diagnosis of Tuberculosis from chest x-rays.

G. Jignesh Chowdary[1,1], Suganya G[2,1], Premalatha M [3,1], Karunamurthy K[4,2]
[1]School of Computer Science and Engineering
[2]School of Mechanical and Building Sciences
Vellore Institute of Technology, Chennai

**Abstract**

Tuberculosis is an infectious disease that is leading to the death of millions of people across the world. The mortality rate of this disease is high in patients suffering from immuno-compromised disorders. The early diagnosis of this disease can save lives and can avoid further complications. But the diagnosis of TB is a very complex task. The standard diagnostic tests still rely on traditional procedures developed in the last century. These procedures are slow and expensive. So this paper presents an automatic approach for the diagnosis of TB from posteroanterior chest x-rays. This is a two-step approach, where in the first step the lung regions are segmented from the chest x-rays using the graph cut method, and then in the second step the transfer learning of VGG16 combined with Bi-directional LSTM is used for extracting high-level discriminative features from the segmented lung regions and then classification is performed using a fully connected layer. The proposed model is evaluated using data from two publicly available databases namely Montgomery Country set and Schezien set. The proposed model achieved accuracy and sensitivity of 97.76%, 97.01% and 96.42%, 94.11% on Schezien and Montgomery county datasets. This model enhanced the diagnostic accuracy of TB by 0.7% and 11.68% on Schezien and Montgomery county datasets.

**Keywords:** Deep learning, transfer learning, VGG16, Tuberculosis, health informatics.

1. Introduction

Tuberculosis(TB) is an infectious disease that is caused due to the bacteria bacillus mycobacterium tuberculosis, which affects the lungs. After HIV, TB is the leading cause of death across the world due to infectious diseases. According to an estimate, nine million new cases of TB are reported each year[1] and out of which 1.2 million people die[2]. TB is a highly contagious disease that spreads through the air when an infected person speaks, sneezes, or coughs. Most of the TB cases are found in southeast Asia and Africa as those people have less resistance to the disease due to malnutrition and poverty. The situation is even worse in patients suffering from immunocompromised disorders like HIV/AIDS[3]. Several antibiotics were available for the treatment of TB. The mortality rate of TB is high when it is not treated, during clinical trials it was reported that the survival chances and cure rates were improved while treating with antibiotics[1]. Unfortunately, the diagnosis of TB is a major challenge in the field of medicine. The definitive test for diagnosing TB is by the detection of Mycobacterium tuberculosis in pus samples or clinical sputum[2][3]. But this test takes several weeks to months to identify this slow-growing bacteria in labs. Another diagnostic test for TB is sputum smear microscopy. This is a hundred-year-old procedure in which the sputum samples are observed under a microscope for the traces of Mycobacterium tuberculosis.

In addition to these, there are immune response-based skin tests that are available to determine whether the person is infected with TB. These tests are reporting promising results but these procedures require advanced equipment which increases the diagnostic cost and also they are time taking. More financial support is required to make these procedures commonplace for TB diagnosis. These tests cannot be used for the rapid screening of huge populations.

Besides these procedures, there is another screening examination based on posteroanterior chest x-rays[4] that can be used for the rapid diagnosis of TB. Chest x-rays provide the thoracic anatomy of the patient. In this procedure, an experienced radiologist is required to manually examine the x-rays for the presence of a TB infection. This examination is highly cost-effective and takes time compared to other tests in TB diagnosis. But there is one single drawback with this examination, which that is they require human intervention, which gives the scope for human error. So to reduce the problem of human error, computerized methods like deep learning and machine learning are used for automatic diagnosis of TB from chest x-rays. These computerized methods have shown promising results in the field of medicine especially for the diagnosis of pneumonia[5, 6, 7], cardiac diseases[8], various types of cancers[9, 10, 11], and brain tumors[13, 12]. An overview of these methods is presented in [14] and [15]. Researchers across the globe developing automatic methods for the diagnosis of TB using these computerized techniques. The recent advancements in the computerized diagnostic methods of TB are presented in [16]. In the study [17], the authors template matching to diagnose miliary TB in an African setting.

With this motivation, in this paper, we propose an automatic approach for the diagnosis of Tuberculosis manifestations in chest x-rays. This is a two-step approach, in the first step the lung regions are segmented from the chest x-rays, and in the second step, the feature extraction and classification are performed. For lung segmentation, we have employed the method presented in [18] and for efficient feature extraction, a transfer learning model with Bi-directional Long Term Short Memory(Bi-LSTM) model is used. And classification is performed using a fully connected layer activated with softmax function. The rest of the paper is organized as follows. Section 2 describes the existing literature, Section 3 describes the proposed methodology, Section 4 discusses the results reported by the proposed model, followed by the conclusion in Section 5.

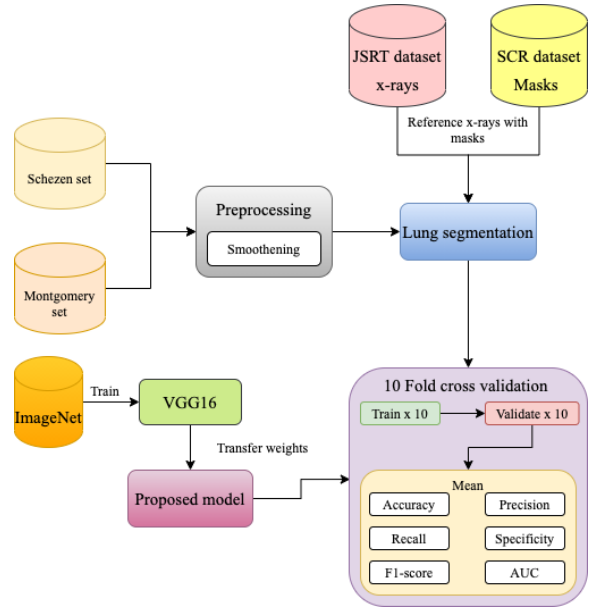

**Figure 1.** Structure of the proposed work.

## 2. Related works

The introduction of digital chest x-rays and the development of digital image processing have given more scope for computer-aided diagnosis(CAD) and screening. Even though the omnipresence of chest x-ray across various specialties of medicine, still chest x-rays remain a complex imaging tool. For a decade several papers were proposed on using computerized methods to diagnose diseases from chest x-rays. However, there is a need for an ample amount of research that is needed to be done for making these systems deployable in real life. According to a survey presented in [19], Van Ginneken et al stated that even after 45 years of research in computerized methods in chest radiology there were no systems that could accurately read chest radiograms. Automated diagnosis of lung nodules using computerized methods is becoming a mature application of decision support for chest x-rays. Several studies were proposed in investigating the capability of CAD systems in the diagnosis of lung nodules[20, 21, 22]. These CAD systems can be used successfully in assisting radiologists in the diagnosis of lung cancer[15]. But nodules are only one of the manifestations of TB.

In recent years, by understanding the complexity in designing a single CAD system for the diagnosis of all problems from x-rays, researchers have focussed on developing systems for specific diseases like pneumonia, tuberculosis, cancer, etc. For designing systems to diagnose lung-related diseases from chest x-rays, it is essential to properly segment the lungs. Even segmentation of other parts like clavicles, heart, and ribs may also be helpful[17]. In [17, 23] the authors performed a comparative study of various segmentation methods like pixel classification, rule-based methods, and active shapes. Their study reported that the pixel classification method performed better than other studied methods. Dawoud[24] used an iterative method to combine shape priors with intensity information. This work was evaluated on the JSRT dataset. In the study [25], the authors used a patch-based feature extraction approach where they have divided the lungs into overlapping regions of different sizes for feature extraction. In the latter study, the authors used the difference between the corresponding regions of the right and left lungs as features. For training this model a separate train set is constructed for each region and the classification is performed using the weighted voting classifier.

Several CAD systems were proposed for diagnosing lung abnormalities from chest x-ray rather than focussing on a specific disease. Among them, there were very few CAD systems specializing in TB diagnosis. Hongeweg et al [26] designed a system by combining the stages of clavicle detection with texture-based abnormal detection to reduce the false positive predictions. The same group of researchers in [26] proposed a clavicle segmentation model built using active shape models and pixel classifiers. The diagnosis of TB from clavicle regions is very difficult because these regions can conceal the TB manifestations in the apex of the lung. In a study performed by Freedman et al [28], it was reported that the suppression of clavicles and ribs can enhance the performance of radiologists in TB diagnosis. The upper lung regions with cavities containing high air-fluid levels indicate that TB has progressed into a highly infectious state. In the study [29] Shen et al proposed a Bayesian approach to detect these cavities in chest x-rays. Xu et al [30] proposed another method that uses the model-based-template-matching technique for the same problem. In this work, the Hessian matrix was used for image enhancement. Santosh et al [31] used the pyramid-HOG(Histogram of Oriented Gradients) method for encoding thoracic edge map features and used MLP(Multi-Layer Perceptron) classifier for TB diagnosis. In the study [32], the authors extracted the texture, edge, shape, and symmetrical features of the lung and performed classification using a voting ensemble of MLP, Random Forest(RF), and Bayesian network(BN). Vajde et al [33] considered three feature sets and used wrapper-based feature selection techniques to select the best feature set to train an artificial neural network algorithm for diagnosis. In the study [33] the authors used Grey-level-Co-occurrence-matrix features, First-Order-Statistical features, and shape-based features to train the support vector machine(SVM) algorithm for diagnosis. Jaeger et al [34] used Frangi, HOG, Local Binary Patterns(LBP), and GIST methods for extracting features and used SVM for classification.

Transfer learning is a learning approach used to enhance the performance of the model while using small datasets. The transfer learning approach works by applying the knowledge that is gained while solving one specific task to be used to solve another similar task. This learning approach is widely used across various domains in medical imaging[35]. Tawsifur et-al[36] performed a study on using the transfer learning of pre-trained models namely MobileNet, SqueezeNet, DenseNet201, VGG19, InceptionV3, ChexNet, ResNet101, ResNet50 and ResNet18 for the diagnosis of TB.

3. **Methods**

3.1 **Data preprocessing**

In this phase, the quantum noise from the chest x-rays is removed. This noise is due to the random dispersal of photons on the receptor plate during the process of x-ray imaging[37] There are several algorithms that were proposed for noise removal[38]. But in this work, a guided filter with a kernel mask($k_m$) of 3 x 3 is used. A Guided filter preserves the edges during the process of noise removal. The denoised image is represented in equation 1.

$$D_j = x_k I_j + y_k, \forall j \in k_m$$

In equation 1, $x_k$ and $y_k$ are the coefficients of input image, $I_j$ is the pixel intensity, $k_m$ represents the kernel mask and $D_j$ is the denoised image.

### 3.2 Proposed methodology

The proposed methodology consists of two modules namely the segmentation module and diagnostic module. In the segmentation module, a three-step approach learned from [18] was used to segment the lung regions from the chest x-rays, and in the diagnostic module, a novel CNN-Bi-LSTM transfer learning method was proposed for feature extraction and classification. The lung segmentation module and the diagnostic modules are explained in sub-section 3.2.1 and 3.2.2.

### 3.2.1 Lung Segmentation module

The chest x-rays contain many anatomical structures of ribs, soft tissues, heart apex, mediastinum, and various organs. These structures show ambiguous appearance in the radiographic responses and make the diagnosis of disease manifestation more complex. Segmentation of lung regions from chest x-rays can resolve issues. Several segmentation methods were proposed[39, 40, 41, 42]. In this work, we used a segmentation approach learned from [18]. This is a three-step approach, the first step involves selecting the reference image(with masks) for each input image. The selection of reference images is based on the similarity between the training image(with mask) and the input image. The similarity is computed using the Bhattacharya coefficient and partial random transform. In the second step, Scale Invariant Feature Transform(SIFT) flow and non-rigid registration method are used for creating lung atlas for input x-ray. And in the last step, the graph-cut approach is used for marking and segmenting the lung regions. Figure 2 shows this segmentation approach on a sample image.

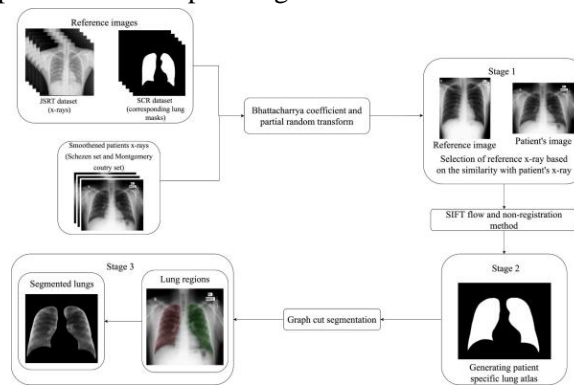

**Figure 2.** Segmentation of a sample x-ray.

For training, chest x-rays from the JSRT dataset are used. The ground truth masks for the images in the JSRT dataset are available in [38]. Note that the lung masks do not contain the posterior inferior regions behind the diaphragm. In this work, we excluded these regions because the manifestation of TB in these regions is less likely to be present.

### 3.2.2 TB diagnostic module

#### 3.2.2.1 CNN based feature extraction

Learning important and highly discriminative features from the input medical images is a complex task in the field of medical imaging. With the rapid improvements in deep learning methodologies, the transfer learning of pre-trained CNN's is extensively used for the extraction of efficient and highly discriminative features. And there are several studies[7, 43, 44] across various image classification applications that have reported excellent results while using this method. With this motivation, we employ a pre-trained CNN named VGG16 developed by Visual Geometry Group(VGG) of Oxford university-trained on ImageNet database for extracting high-level discriminative features.

The VGG16 CNN consists of five convolutional blocks, each block consisting of two or three convolutional layers as shown in figure 3 for feature extraction. In each block, the convolutional layers have the same number of filters and the number of filters becomes double after each max-pooling layer. All the convolutional layers have the same receptive fields of 3 x 3, which results in the increase of non-linearity in the feature extraction module. The max-pooling layers are used in this network for reducing the size of feature maps produced by the convolutional layers by maintaining a pool size of 2 x 2, thus these pool layers reduce the size of the feature map by half in its width and height. The main reason behind such design is to use minimum computational resources for extracting high diverse features. Following the feature extraction module, there is one classification module with three fully connected layers consisting of 512, 256, and 1000 neurons. The last fully connected layer is activated with the softmax function and is used for classification. Except for the last fully connected layer, the remaining convolutional and dense layers are activated with the ReLU function. Since in this work, the VGG16 is used as a feature extractor the classification module of the network is removed and the parameters of the original network are preserved.

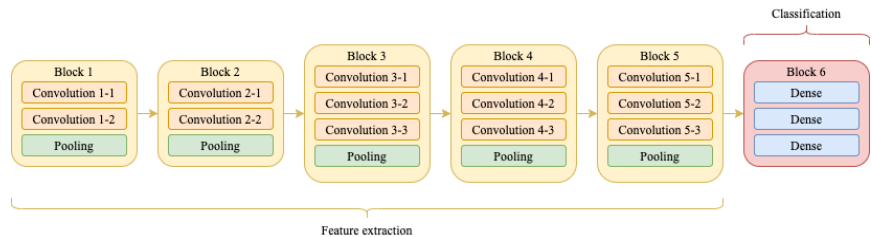

**Figure 3**. Structure of VGG16.

#### 3.2.2.2 Bi-LSTM based class dependency learning

Recurrent neural networks(RNN) is a subset of deep learning mainly used for dealing with sequential data like temporal series and textual data. These RNN's are capable of exploiting implicit dependencies among inputs. RNN's contain recurrent neurons, their activations are dependent on previous hidden states and current input. However, these RNN's suffer from the problem of vanishing gradients and they find it difficult to learn long-term dependencies. In this work, we use the RNN based LSTM approach proposed in [45] to model class dependencies, these have shown excellent performance in [46, 47, 48, 49, 50] for processing long sequences.

LSTM doesn't sum up the inputs like convolutional RNN's, instead, these networks have hidden units known as LSTM units, where the information regarding the class dependencies between different categories is memorized, transmitted, and updated with a memory cell and various gates. But a single LSTM is not sufficient to get a complete view of inter-class relevance, so a comprehensive Bidirectional-LSTM(Bi-LSTM) approach is proposed in this paper. This Bi-LSTM consists of two LSTM placed in opposite directions with each other and the units in this network are not only updated with previous states but also with subsequent ones.

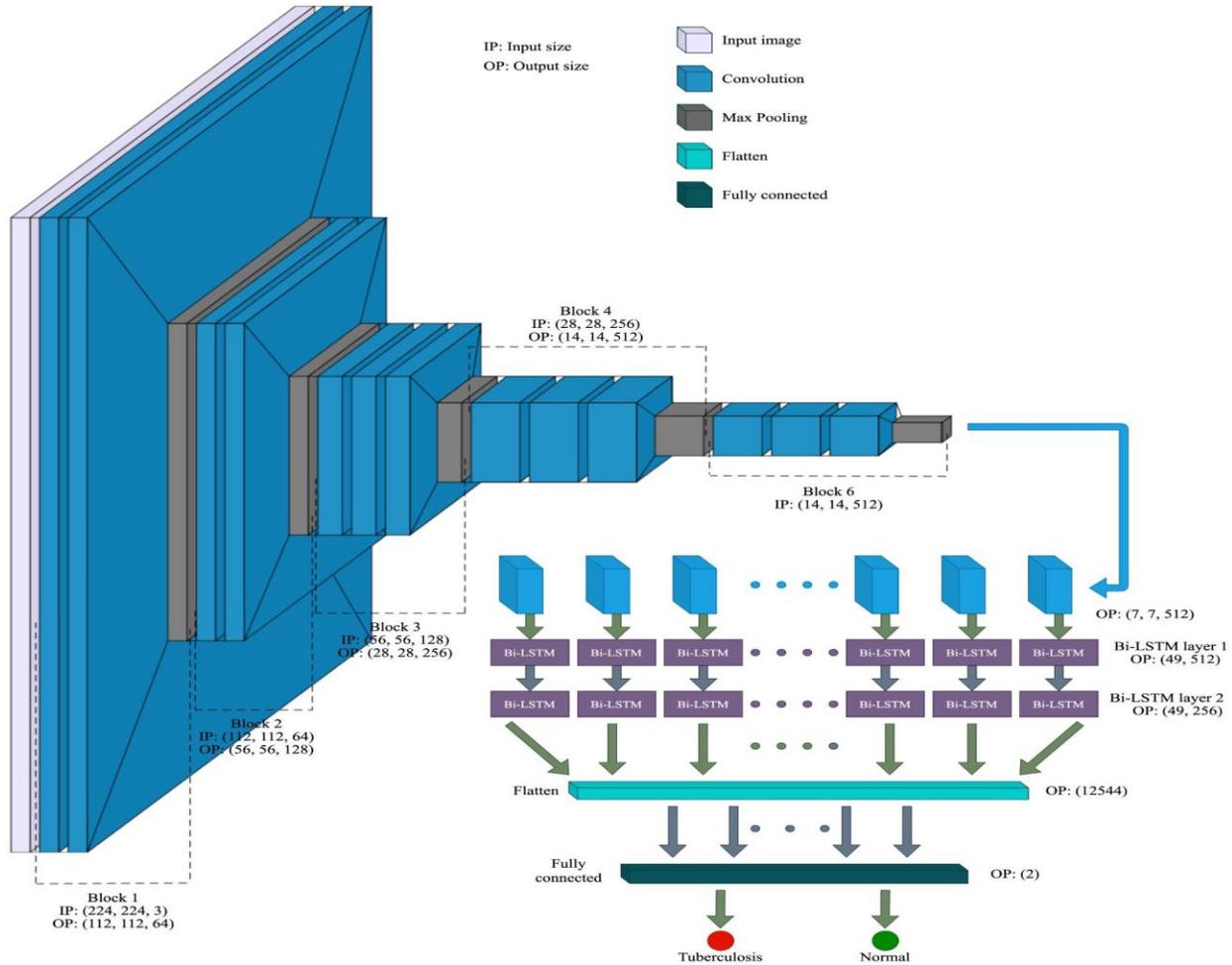

**Figure 4.** Structure of proposed work.

As seen in figure 4, two Bi-LSTM's are used in this work. The output from the CNN feature extractor cannot be given as an input to the Bi-LSTM layers due to shape constraints. So a Reshape layer is used to connect the CNN module with the Bi-LSTM layers. After feature extraction, the extracted features are flattened using a flatten layer and are fed into a dense layer with two neurons activated with a softmax activation function for classification. The summary of the proposed model is shown in table 1.

| Block | Layer | Output shape |
|---|---|---|
| Input | Input layer | (224, 224, 3) |
| Block 1 | Convolution | (224, 224, 64) |
| | Convolution | (224, 224, 64) |
| | Max Pooling | (112, 112, 64) |
| BLock 2 | Convolution | (112, 112, 128) |
| | Convolution | (112, 112, 128) |
| | Max Pooling | (56, 56, 128) |
| Block 3 | Convolution | (56, 56, 256) |
| | Convolution | (56, 56, 256) |
| | Convolution | (56, 56, 256) |
| | Max Pooling | (28, 28, 256) |
| Block 4 | Convolution | (28, 28, 512) |
| | Convolution | (28, 28, 512) |
| | Convolution | (28, 28, 512) |
| | Max Pooling | (14, 14, 512) |
| Block 5 | Convolution | (14, 14, 512) |
| | Convolution | (14, 14, 512) |
| | Convolution | (14, 14, 512) |
| | Max Pooling | (7, 7, 512) |
| Reshape | Reshape | (49, 512) |
| Bi-LSTM block | Bi-LSTM | (49, 512) |
| | Bi-LSTM | (49, 256) |
| Classification | Flatten | (12544) |
| | Dense | (2) |

Table 1. Summary of the proposed model.

## 4. Results and discussion

### 4.1 Databases

For our work, three chest x-ray datasets are used. The first dataset is used for training our lung segmentation model and the second and third dataset are used for training and testing our classifier.

Our first dataset is from the Japanese Society of Radiological Technology(JSRT). The data in this dataset is the result of a study performed for understanding the performance of radiologists in identifying pulmonary modules[51]. This dataset contains 247 chest x-rays each of size 2048 x 2048 pixels in a 12-bit grayscale color coding scheme extracted from 14 medical centers. Out of these 247 chest x-rays, 154 chest x-rays are abnormal and 93 chest x-rays are normal. Each abnormal chest x-ray has one pulmonary nodule, and these nodules are categorized into five classes ranging from obvious to extremely subtle. These lung nodules don't affect the shape of the lung. So we can use the entire dataset for training our segmentation model. In order to do so, we used segmentation masks provided by Ginneken et al [52] in their SCR dataset (Segmentation in Chest Radiographs). This dataset contains manually segmented masks for each chest x-ray in the JSRT dataset. Few samples from both datasets are shown in figure 5.

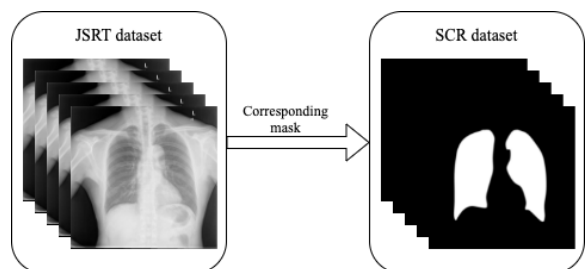

**Figure 5.** Sample x-rays and corresponding masks from JSRT dataset and SCR dataset.

Our second dataset, the Shenzhen dataset(Shenzhen set)[53], is developed by one of the best hospitals for treating infectious diseases in China named Shenzhen hospital located in the Guangdong province. All the images were collected in September 2012 by the hospital radiology department. This dataset contains a total of 662 chest x-rays, out of which 336 chest x-rays with TB infection and 326 chest x-rays are normal. These x-rays are extracted using Philips DR digital diagnostic system.

Our third dataset, the Montgomery Country(Montgomery set) dataset is a subset of the large chest x-ray database collected during the tuberculosis control program organized by the Department of Health and human services of the Montgomery Country of Maryland[54]. This dataset contains a total of 138 posteroanterior chest x-rays out of which 58 chest x-rays contain lungs with tuberculosis manifestations and 80 chest x-rays are normal. All the images in this dataset are 12-bit grayscale color coding schemes and are taken using Eureka stationary x-ray devices. Few samples of x-rays from Shenzhen and Montgomery datasets are shown in figure 6.

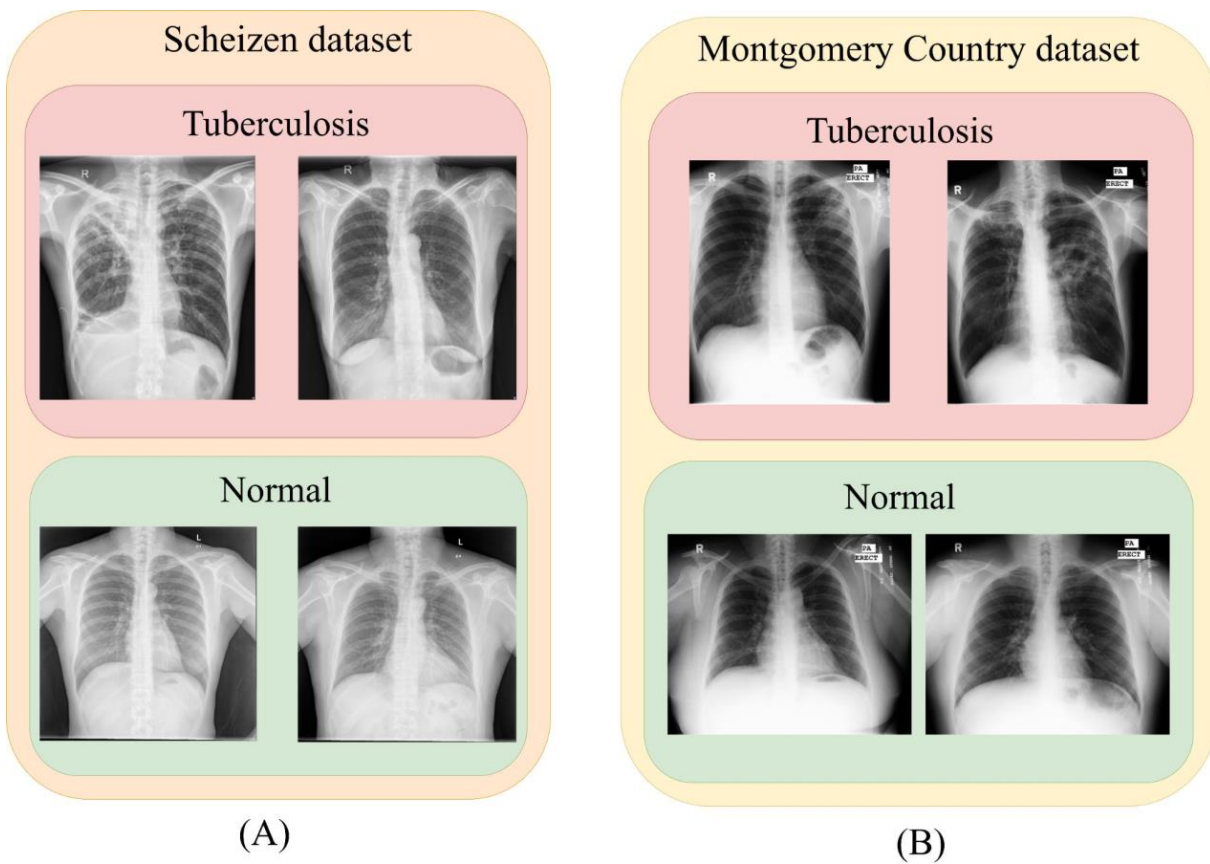

**Figure 6.** Sample x-rays from (A) Shenzhen and (B) Montgomery country datasets.

### 4.2 Performance metrics

The robustness of the proposed model in diagnosing TB from chest x-rays is evaluated using six performance metrics namely accuracy, sensitivity, specificity, precision, Negative Predicted Value(NPV), and F1-Measure. These performance metrics were calculated using True Positive(TP), True Negative(TN),

False Positive(FP), and False Negative(FN) values. These performance metrics are represented in equations 1-6.

$$Accuracy = \frac{(TP + TN)}{(TP + TN + FP + FN)}$$

$$Precision = \frac{TP}{(TP + FP)}$$

$$Recall = \frac{TP}{(TP + FN)}$$

$$Specificity = \frac{TN}{(TN + FP)}$$

$$F1 - score = \frac{(2 \times precision \times recall)}{(precision + recall)}$$

In the above equations, TP represents the number of samples that were correctly predicted as TB, TN represents the number of samples that were correctly predicted as normal, FP and FN represent the number of samples that were wrongly predicted as TB and normal. In addition to these metrics Area Under the Curve(AUC) is also considered for understanding the efficacy of the proposed model.

### 4.3 Results reported

In this work, the ten-fold cross-validation protocol is employed for validating the performance of the model. The ten-fold cross-validation results reported by the proposed model on Shenzhen and Montgomery datasets are shown in table 2.

| Dataset | Accuracy | Precision | Recall | Specificity | F1-score | AUC |
|---|---|---|---|---|---|---|
| Shenzhen set | 97.76 | 98.48 | 97.01 | 98.50 | 97.74 | 0.9786 |
| Montgomery set | 96.42 | 100 | 94.11 | 100 | 96.96 | 0.9589 |

**Table 2.** 10-fold cross-validation results reported by the proposed model on Shenzhen set and Montgomery set.

### 4.4 Comparison

#### 4.4.1 Comparison with pre-trained models

In this work, the ten-fold cross-validation results reported by the proposed model are compared with the performance of well-known transfer learning models namely InceptionV3, VGG16, VGG19, ResNet50, Xception, DenseNet201, MobileNet, MobileNetV2, and InceptionResNetV2. For making these models suitable for our task, the last fully connected layer is modified by changing the number of units to 2. Then these models are trained with data from both the datasets and their performance is shown in tables 3 and 4.

| Models | Accuracy | Precision | Recall | Specificity | F1-score | AUC |
|---|---|---|---|---|---|---|
| InceptionV3 | 85.07 | 77.27 | 91.07 | 80.76 | 83.6 | 0.8495 |
| VGG16 | 85.07 | 83.33 | 85.93 | 84.28 | 84.61 | 0.8504 |
| VGG19 | 86.56 | 93.93 | 81.57 | 93.1 | 87.32 | 0.8667 |
| ResNet50 | 85.82 | 96.96 | 79.01 | 96.22 | 87.07 | 0.8598 |
| Xception | 79.1 | 98.48 | 70.65 | 97.61 | 82.27 | 0.7938 |
| DenseNet201 | 84.32 | 96.96 | 77.1 | 96.07 | 85.9 | 0.8451 |
| MobileNet | 84.32 | 71.21 | 95.91 | 77.64 | 81.73 | 0.8413 |
| MobileNetV2 | 76.11 | 53.03 | 96.22 | 68.36 | 68.62 | 0.7577 |
| InceptionResNetV2 | 88.8 | 83.33 | 93.22 | 85.33 | 88 | 0.8872 |
| **Proposed model** | **97.76** | **98.48** | **97.01** | **98.50** | **97.74** | **0.9786** |

**Table 3.** Performance comparison on Shenzhen set.

| Models | Accuracy | Precision | Recall | Specificity | F1-score | AUC |
|---|---|---|---|---|---|---|
| InceptionV3 | 80 | 88.88 | 72.72 | 88.88 | 79.99 | 0.8333 |
| VGG16 | 70 | 87.5 | 58.33 | 87.5 | 70 | 0.7291 |
| VGG19 | 85 | 75 | 85.71 | 84.61 | 79.99 | 0.8334 |
| ResNet50 | 90 | 87.5 | 87.5 | 91.66 | 87.5 | 0.8958 |
| Xception | 80 | 62.5 | 83.33 | 78.57 | 71.42 | 0.8125 |
| DenseNet201 | 70 | 87.5 | 58.33 | 87.5 | 70 | 0.7291 |
| MobileNet | 80 | 87.5 | 70 | 90 | 77.77 | 0.8125 |
| MobileNetV2 | 80 | 77.77 | 77.77 | 81.81 | 77.77 | 0.8129 |
| InceptionResNetV2 | 80 | 62.5 | 83.33 | 78.57 | 71.42 | 0.7708 |
| **Proposed model** | **96.42** | **100** | **94.11** | **100** | **96.96** | **0.9589** |

**Table 4.** Performance comparison on Montgomery set.

In medical imaging, accuracy and recall are the two performance metrics that are used to determine the robustness and efficiency of the model. On the Shenzhen set, InceptionResNetV2 and MobileNet are the models with the highest accuracy and recall among the considered networks. On the Montgomery set, ResNet50 is the model with the highest accuracy and recall among the other networks. The proposed model reported an enhanced performance in diagnosing TB from both datasets. Table 4 shows the percentage increase in accuracy and recall reported by the proposed model than the high performing models on both datasets is shown in table 5.

| Datasets | High performing models | | | Proposed model | | Percentage increase | |
|---|---|---|---|---|---|---|---|
| | Models | Accuracy | Recall | Accuracy | Recall | Accuracy | Recall |
| Shenzhen Set | Inception ResNetV2 | 88.8 | 93.22 | 97.76 | 97.01 | 8.96 | 3.79 |
| | MobileNet | 76.11 | 96.22 | | | 21.65 | 0.79 |
| Montgomery set | ResNet50 | 90 | 87.5 | 96.42 | 94.11 | 6.42 | 6.61 |

**Table 5.** Percentage increase in accuracy and recall while using the proposed model than other pretrained networks.

### 4.4.2 Comparison with other existing literature

In this work, the performance of the proposed model is compared with other existing works discussed in the literature review section of the paper. The cross-validation accuracy of the proposed model is higher than the accuracy reported by the works. This comparison of accuracy on Montgomery and Shenzhen sets were shown in figure 7.

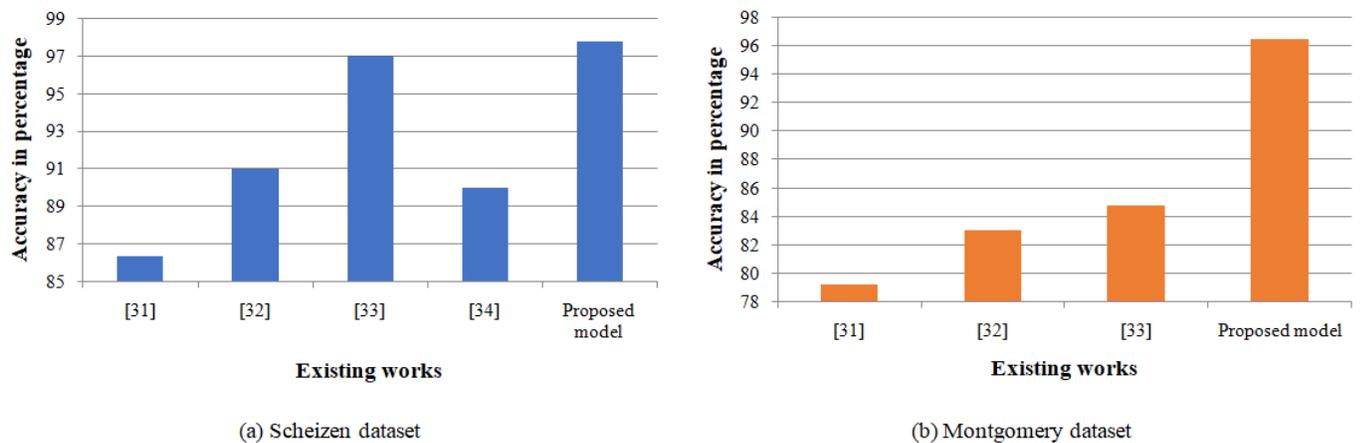

(a) Scheizen dataset    (b) Montgomery dataset

**Figure 7.** Performance comparison on Scheizen and Montgomery datasets.

From figure 5, the methods proposed in [33] reported high accuracy of 97.03% and 84.75% on Shenzhen set and Montgomery datasets other than the proposed model. This clearly illustrates that while using the proposed model the accuracy in diagnosing TB is increased by 0.7% and 11.68% on the Shenzhen and Montgomery datasets. This percentage increase from table 5 and figure 7 shows that the proposed model is efficient and effective than other pre-trained models and existing works.

**Conclusion**

In this paper, an automatic approach for the diagnosis of Tuberculosis from chest x-rays is presented. This approach focuses on diagnosing TB from segmented lung regions. Since the work focuses more on diagnosis rather than segmentation, we used a graph cut-based segmentation method presented in [18] for extracting the lung regions from chest x-rays. For the diagnosis, we present an efficient feature extraction module built by combining a transfer learning model with a recurrent neural network. In this feature extraction module, the VGG16 pre-trained network is used for extracting deep features and two bi-directional LSTM are used for extracting the class-dependent features. The proposed model is evaluated on Scheizen and Montgomery County datasets. The proposed model reported higher performance than other pre-trained networks and existing works by achieving a sensitivity of 97.01% and 94.11% on Scheizen and Montgomery Country datasets.